# Influence of Platinum Thin Films on the Photophysical and Quantum Properties of Near-Surface NV Centers

Joachim P. Leibold[1,5,6,†], Lina M. Todenhagen[2,5,6,†], Matthias Althammer[3,5,6], Nikhita Khera[4], Elke Neu[4], Martin S. Brandt[2,5,6], Hans Huebl[3,5,6], Dominik B. Bucher[1,6]*

[1] Department of Chemistry, School of Natural Sciences, Technical University of Munich, Garching, Germany
[2] Walter Schottky Institute, Technical University of Munich, Garching, Germany
[3] Walther-Meißner-Institute, Bavarian Academy of Sciences and Humanities, Garching, Germany
[4] Department of Physics and Research Center OPTIMAS, RPTU Kaiserslautern Landau, Kaiserslautern, Germany
[5] Department of Physics, School of Natural Sciences, Technical University of Munich, Garching, Germany
[6] Munich Center for Quantum Science and Technology (MCQST), Munich, Germany

*Corresponding author: dominik.bucher@tum.de
†These authors contributed equally to this work

**Abstract**

**Nitrogen-vacancy (NV) centers in diamond are optically addressable spin defects with great potential for nanoscale quantum sensing. A key application of NV centers is the detection of external spins at the diamond surface. Among metals, platinum thin films – widely used in spintronics, catalysis and electrochemistry – provide a particularly interesting system for such studies. However, the interaction between NV centers and metals is known to affect their quantum sensing capabilities. In this work, we study five platinum-covered diamond samples containing shallow NVs created via nitrogen implantation with different energies (2.5-60 keV) and investigate the optical and quantum properties of NV ensembles beneath the metal films. We find a substantial reduction of the photoluminescence lifetime and a pronounced decrease of the NV⁻ population for NV ensembles located near the platinum layer. As a result, optically detected magnetic resonance experiments could only be efficiently performed on diamonds implanted with at least 20 keV, where we observed a strong increase in the $T_2$ coherence time beneath the platinum thin films. Our study describes the various processes affecting NV centers near platinum films and provides guidance for the integration of thin metal films with near-surface NV centers.**

## 1. Introduction

The nitrogen-vacancy (NV) center in diamond has attracted increased interest as a solid-state spin system for quantum technology applications in the last years [1–4]. It is composed of a nitrogen atom and an adjacent vacancy occupying two neighboring carbon lattice sites. Its negative charge state shows spin-dependent photoluminescence (PL), allowing for easy optical access, initialization and readout of the system [5,6]. NV centers have been applied to measure strain, temperature, electric fields, and, in particular, magnetic fields [7–10]. Microwave-controlled pulse sequences enable the detection of magnetic signals over a wide frequency range [11–13]. One key advantage of the NV center is that it can be created just a few nanometers beneath the diamond surface and thus be in proximity to samples deposited on top. This property allows to realize surface-sensitive detection schemes of nanoscale samples or even single spins via the observation of magnetic resonance signals [14–21]. However, near-surface NV centers show significantly deteriorated properties compared to those in the bulk [22–26]. This can e.g. be related to noise sources specific to the surface [27] or the surface termination [26,28,29] that impacts the charge state of the NV center. In most applications, a highly oxidized diamond surface is prepared to stabilize the negative charge state of the NV center [27]. In many cases, however, the sample of interest needs to be deposited onto the diamond surface [18,21,30–34]. In particular, conducting samples can strongly interact with near-surface NV centers, resulting in PL quenching or reduced charge state stability, which significantly reduces the NV sensing capabilities [33,35–37]. For these materials, a balance must be found between placing

NV centers close enough to the sample to enhance interaction strength and keeping a sufficient distance to minimize the negative effects. One highly interesting application is the integration of platinum (Pt) thin- and ultrathin-films with near-surface NV centers, as platinum plays an important role in catalysis or electrochemistry. With the capabilities of NV centers in detecting nanoscale NMR signals, probing $^{195}$Pt would offer insights into the structure of surface-bound catalysts or the electrochemical interface [38–40]. Similarly, NV NMR could be used to intrinsically probe the spin-Hall physics in thin platinum films [41–44]. Moreover, in recent years, also the electrical readout of the NV spin state gained increasing attention due to its suitability for device integration and miniaturization [45–47]. This alternative approach relies on measuring an NV-related photocurrent and typically employs metal electrodes deposited directly on the diamond surface, possibly affecting the NV centers.

Here, we investigate the impact of metal thin films on ensembles of near-surface NV centers using platinum as a particularly interesting example. For that, we prepared different diamond samples with varying NV depth distributions, partly covered with a ~ 5 nm platinum film (see Figure 1 (a)). We then examine key properties relevant to NV-based sensing applications, including the NV centers' charge state, the NV centers' PL lifetime, and their spin properties (Figure 1 (b)). We find that the platinum thin films reduce drastically the population of the NV centers in the negative charge state and their PL lifetime. The strength of these effects increases with the proximity of the NV centers to the surface. Notably, for the shallowest ensemble where advanced NV measurement protocols could be performed, the spin properties (in particular the coherence time $T_2$) improve on the platinum-covered side with respect to the bare reference. Similar phenomena were also observed by Henshaw et al. for implanted NV centers beneath copper thin films [37], indicating that these effects could be very general consequences of the proximity to metallized interfaces. These findings highlight that the influence of metal thin films on near-surface NV centers needs to be carefully considered in nanoscale sensing as well as in electric readout scenarios.

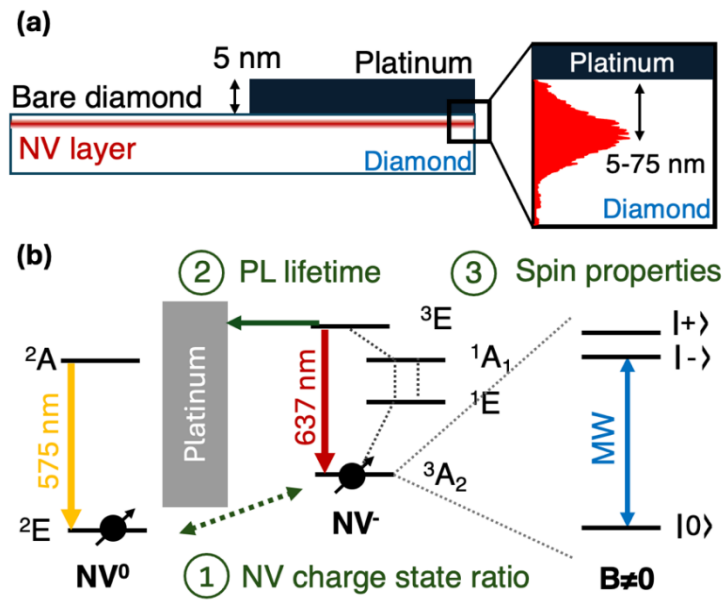

**Figure 1. Platinum thin films on a diamond chip with near-surface NV centers. (a)** The diamond samples contain near-surface layers of NV centers created at different depths (5-75 nm) using varying nitrogen implantation energies. Half of the diamond surface was covered with a 5-nm-thick platinum film, whereas the second half remains a bare diamond surface. This part of the diamond acts as a reference. **(b)** Possible effects of platinum films on near-surface NV centers include: (1) The metal coating can induce surface band bending and charge transfer, changing the NV charge state. (2) The metal can introduce additional non-radiative decay channels similar to the Förster mechanism and reduce the NV centers' photoluminescence (PL). (3) Finally, the metal-diamond interface can influence also the spin properties of the NV center. These are measured in a subset of the NV- states that is defined by a resonant microwave (MW) transition.

## 2. Experimental Section

### 2.1 Sample Preparation

All experiments have been performed on diamond chips containing ensembles of near-surface NV centers. The high-purity diamond samples (electronic grade, grown via chemical vapor deposition by element six) were implanted with nitrogen ions ($^{14}$N) using five kinetic energies from 2.5 keV to 60 keV and a fluence of $2 \times 10^{12}$ ions/cm$^2$. Implantation at an angle of 7° with respect to the diamond surface was carried out by Coherent Corp. (formerly Innovion). Subsequent vacuum annealing of the diamonds in an in-house-built furnace leads to the formation of NV centers, which has been described in detail before [48]. Prior to platinum deposition (see Section 2.3), we exposed the diamond surface to triacid cleaning to stabilize the negative NV charge state via oxygen termination [48].

### 2.2 SRIM Simulations

To estimate the depth distribution of the NV centers in our ensembles, we performed Monte-Carlo-based implantation depth simulations using the Stopping and Range of Ions in Matter (SRIM) software [49]. In this simulation, a 200-nm-thick carbon target layer with a density of 3.5 g cm$^{-3}$ was implanted at an angle of 7° using $^{14}$N nitrogen atoms. We simulated the same configurations for the five different implantation energies used in this study. This results in distributions of stopping depths, from which we used the mean distance (with respect to the diamond surface) as depth parameter for later analysis. The SRIM simulation results are summarized in Table 1. We note that experiments have shown that SRIM simulations generally tend to underestimate the actual implantation depths [50,51].

**Table 1.** Overview of the mean depth and the longitudinal standard deviation (straggle) of the implanted nitrogen atoms, which are obtained from the SRIM simulations as summarized in Section 2.2.

| Implantation energy | 2.5 keV | 5 keV | 10 keV | 20 keV | 60 keV |
|---|---|---|---|---|---|
| Mean depth | 4.6 nm | 8.0 nm | 14.7 nm | 27.2 nm | 73.7 nm |
| Straggle | 1.8 nm | 3.1 nm | 5.3 nm | 8.5 nm | 16.2 nm |

### 2.3 Platinum deposition

The platinum thin films were deposited on the cleaned diamond surface using a custom-built ultra-high vacuum sputtering system (Bestec GmbH) with a base pressure of $3 \times 10^{-9}$ mbar in the deposition chamber. For the sputtering, a platinum target with a purity of 99.99% was used. Sputtering at room temperature in $5 \times 10^{-3}$ mbar argon (purity 99.9999%) at a DC power of 7 W for 150 seconds led to film thicknesses of 5 nm. The resulting film thickness has been calibrated from previous deposition runs for other experiments [52,53] via x-ray reflectivity and atomic force microscopy. Both methods yield a typical root-mean-square surface roughness of 0.3 nm for these films. One half of each diamond sample was protected by a Kapton foil during Pt deposition to obtain a bare diamond side used for internal referencing.

### 2.4 Photoluminescence measurements

The PL spectra shown in Figure 2 were recorded with a custom-built confocal Raman spectrometer. For optical excitation, we used a 532 nm laser (Laser Quantum, torus 532), focused on the NV layer with a microscope objective (Olympus, UMPlanFl, 50x, numerical aperture 0.8). Note, that excitation and PL collection occurred through the back side of the diamond to avoid light-screening by the platinum film. The collected light is filtered with a 532 nm notch filter and a 532 nm long-pass filter. The light is then guided into a commercial spectrometer (HORIBA, iHR550). We used a grating with 300 lines/mm and adjusted slit width and integration time for the different PL intensities of the samples. The relative NV charge state populations are calculated from the integrated zero phonon lines of the neutral (575 nm) and negative (637 nm) charge state in the emission spectrum [54,55]. The NV⁻ population is normalized to the total NV concentration according to [NV⁻]/([NV⁻]+[NV⁰]). For the neutral charge state, we apply a correction factor of 1.7 to account for the different Debye-Waller factors and photoluminescence excitation efficiencies of NV⁰ and NV⁻ at 532 nm [55,56].

## 2.5 Fluorescence lifetime measurements

The fluorescence lifetime measurements shown in Figure 3 were performed with a custom-built confocal scanning microscope (numerical aperture 0.8, pinhole size 50 μm). In the experiment, the NV centers were excited by short laser pulses from a tunable (400–840 nm) pulsed supercontinuum laser (NKT SuperK EXW-12, pulse length ~ 50 ps). We filtered the emission of the laser to select a 10 nm-wide wavelength window centered around 532 nm. The collected PL signal was sent to a single-photon counter (Excelitas SPCM-AQRH-14, quantum efficiency ~ 70%). For the time-resolved analysis of the NV⁻ PL, we used correlation electronics (PicoQuant, PicoHarp 300). We fit the measured decay of PL intensity $I(t)$ over time $t$ to the stretched exponential

$$I(t) = I_0 \cdot e^{-\left(\frac{t}{t_{PL}}\right)^s} + I_{off} \qquad (1)$$

where $I_0$ and $I_{off}$ denote the intensities of initial and background PL, respectively, $s$ is the stretching parameter and $t_{PL}$ the characteristic timescale of the decay. Additionally, Equation (1) is corrected by a second exponential summand $I_{0b} \cdot e^{-\left(\frac{t}{2.4\,\text{ns}}\right)}$ to account for the finite temporal resolution of the measurement setup (instrument respond function, IRF). Using Equation (1), we obtain the PL lifetimes for emitters on the Pt-coated side $t_{PL,\,Pt}$ and the bare reference side $t_{PL,bare}$ and assume the relative lifetime to depend on the distance $d$ between the NV ensemble and the platinum film as

$$\frac{t_{PL,Pt}}{t_{PL,bare}}(d) = \left(1 - \frac{a}{1+\left(\frac{d}{d_0}\right)^p}\right) \qquad (2)$$

This relationship is based on a reduction in $t_{PL}$ due to energy transfer from the NV emitters to acceptor states in the platinum layer with a mechanism similar to FRET or MIET. $a$ denotes a proportionality constant to describe the correlation between energy transfer efficiency and actual lifetime reduction. The exponent $p$ depends on the nature of the energy accepting states in the platinum layer and $d_0$ is the length scale of the interaction [57,58].

## 2.6 Optically detected magnetic resonance setup

The setup used for the optically detected magnetic resonance experiments (ODMR) depicted in Figure 4 has been described in detail before [18,48,59]. To excite and polarize the NV ensembles, we used a 532 nm laser (Coherent Cooperation, Verdi G8). For pulsed laser experiments, the beam was sent through an acousto-optic modulator (Gooch & Housego, 3260-220) and then focused on the diamond sample from the backside under a total internal reflection geometry [18,48]. The emitted NV fluorescence response was collected and guided onto an avalanche photodiode (APD, Laser Components GmbH, ACUBES3000-10). To block green stray light, a 647 nm long-pass filter was inserted in front of the APD. For microwave (MW) pulses, the output of a MW source (Windfreak Technologies, SynthHD) is connected to a 90° phase splitter (Mini-Circuits, ZX10Q-2-27-S+), whose output ports are TTL-controlled with MW switches (Mini-Circuits, ZASWA-2-50DRA+), combined (Mini-Circuits, ZX10-2–442-S+) and then fed into a MW amplifier (Mini-Circuits, ZHL-16W-43-S+). The output of the amplifier was connected to a short-circuited coaxial cable, which serves as loop antenna for NV⁻ control [48]. The readout of the APD voltage is recorded with a data acquisition device (National Instruments, USB-6229 DAQ). The laser, MW pulses, and data acquisition were controlled using trigger electronics (SpinCore Technologies, PulseBlaster ESR-500 Pro). The magnetic field was aligned with one of the four principal directions of the NV center. The experiments have been conducted at an NV⁻ resonance frequency of ~ 2 GHz, corresponding to a magnetic field of ~ 310 Gauss. The polarization of the excitation light was optimized to the chosen NV⁻ orientation with a λ/2 waveplate.

### 2.7 Pulse sequences

All NV⁻ spin control experiments (Rabi oscillation, relaxation, decoherence) were performed with common pulse sequences described previously in literature [12,48]. Within all pulse sequences, two readouts, signal and reference, were recorded to cancel out low frequency (< 50 kHz) noise, in particular the noise arising from laser amplitude fluctuations [48]. For the Rabi sequence, the applied MW pulse duration was increased stepwise at a constant output power. Typically, we achieved π-pulse times of ~ 50 ns for the experiments in this work. For the PL intensity measurements in Figure 4 (b), the APD voltage was recorded for both the platinum-covered and bare diamond surface. The value on the Pt-covered side was normalized to the APD voltage recorded on the bare surface. For the $T_1$ relaxation measurements, a π-pulse was applied to invert the polarization of the NV ensemble and the decay to the equilibrium is monitored. The data was then fitted to a single exponential decay, corresponding to $s = 1$ in Equation (1). To measure the decoherence time $T_2$, we used the Hahn echo sequence, where we applied a π-pulse centered between an initialization and a projection π/2-pulse. The obtained decay signal was fitted to a stretched exponential function with a common stretch parameter $s=0.7$ in Equation (1). The full data contain an oscillating part due to interactions with ¹³C nuclei that is not considered for the decoherence time, which is given by the envelope of this modulation [60].

## 3. Results

### 3.1 NV charge state ratio

To characterize the optical properties of the NV ensembles, we first recorded the PL spectra (Figure 2) of our five diamond samples under continuous 532 nm illumination with 1 mW optical power. For the bare diamonds in Figure 2 (a), the PL intensity increases with increasing implantation energy, which is a consequence of the larger number of vacancies generated per implanted nitrogen atom and thus higher NV formation yield [61]. For the diamonds with platinum films, Figure 2 (b) reveals a significant reduction in the total PL compared to the bare diamonds, with the effect being more pronounced at lower implantation energies. For the two samples with the lowest implantation energy, the PL below the platinum vanishes completely.

To investigate the effect of platinum coatings on the NV charge state, we characterize the relative population of NV⁻ and NV⁰ using their respective PL spectra as described in the Experimental Section. The charge state below the platinum film could not be determined for the two lowest implantation energies due to the strongly reduced PL (Figure 2 (b)). Consequently, these data points were set to zero for further analysis and possible explanations for the total quenching of NV⁻ PL, including a conversion to the positively charged NV⁺, are discussed in Section 4. Figure 2 (c) shows the percentage of negatively charged NV centers for the platinum-covered (NV⁻$_{Pt}$) and the bare diamond sides (NV⁻$_{Bare}$). Notably, even in the bare diamonds the NV charge state differs between implantation energies, which is most likely a consequence of implantation energy-dependent concentrations and distributions of both nitrogen donors and vacancy-related acceptor states created in the implantation process. To filter out this effect and reliably assess changes in the charge state population induced by the Pt film, we therefore calculate the ratio of the NV⁻ populations below the platinum-covered surface with respect to the bare diamond. This ratio is plotted against the simulated mean implantation depth (see Experimental Section, Table 1) in Figure 2 (d). For all implantation depths, we observe a significant shift towards the neutral NV⁰, that becomes stronger with decreasing distance to the Pt film.

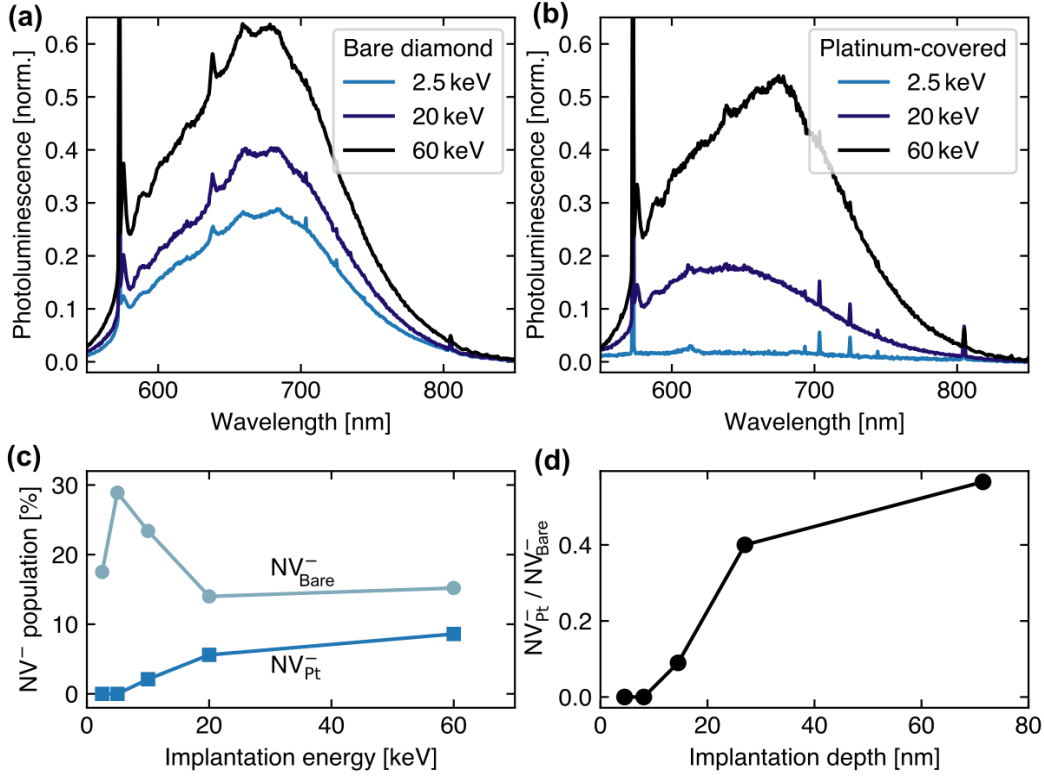

**Figure 2. NV photoluminescence and charge state.** (a) PL spectra of NV ensembles on the bare diamond side. (b) PL spectra on the platinum-covered side. All spectra are normalized to the first order diamond Raman peak. (c) NV⁻ populations with respect to total NV population for the five samples with different implantation energies for bare (NV⁻$_{Bare}$) and the platinum-covered diamond (NV⁻$_{Pt}$). The populations are obtained by integrating the respective ZPLs (see Experimental Section). (d) Ratio of the NV⁻ populations on the platinum-covered side with respect to the bare diamond as a function of the implantation depths simulated via SRIM (see Experimental Section, Table 1).

### 3.2 NV⁻ photoluminescence lifetime

It is well known that emitters near metallic interfaces often experience PL quenching and a reduction in excited state lifetime caused by non-radiative decay channels induced by the metal. This phenomenon is called metal-induced energy transfer (MIET) [58] or surface energy transfer (SET) [62] and is strongly related to the Förster resonant energy transfer (FRET) for molecular systems [57]. A similar mechanism has been previously observed for NVs interfaced with 2-dimensional materials like graphene and WSe₂, as well as with copper and other metals [33,35,37,63–68].

For that reason, we determined the PL lifetime of the NV⁻ charge state from decay measurements as shown for the example of the 20 keV diamond sample in Figure 3 (a). The PL lifetimes for all our five diamond samples are plotted in Figure 3 (b) as a function of the implantation energy. In accordance with literature, PL lifetimes of NV centers in the bare diamond $t_{PL,bare}$ are increasing towards the surface. This is a consequence of the decreased number of states in the air environment, into which the excited NV state can decay [69–71]. The platinum film on the other hand reduces the PL lifetime $t_{PL,Pt}$, with the strongest effect on the NV centers with the lowest implantation energies. The lifetimes of the three NV ensembles closest to the platinum film are highly reduced, while the lifetime recovers to the bare reference diamond for the deepest ensemble with an implant energy of 60 keV. We quantify the reduction of the NV⁻ PL lifetime on the platinum-covered side relative to its counterpart below the bare surface. Figure 3 (c) shows this relative lifetime plotted over the mean implantation depth obtained from the SRIM simulations (see Experimental Section, Table 1).

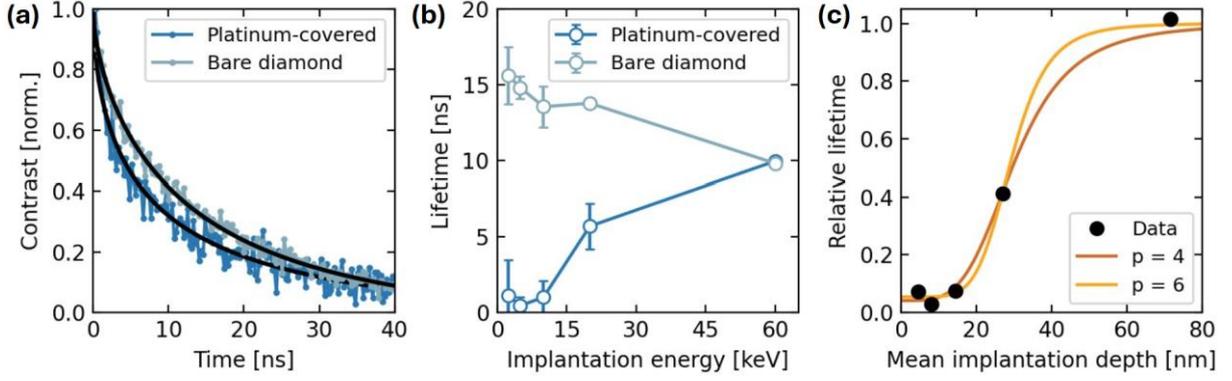

**Figure 3. NV⁻-ensemble photoluminescence lifetimes. (a)** PL lifetimes for the 20 keV diamond sample below the platinum thin film (dark blue) and on the bare reference surface (light blue). Black lines are stretched exponential fits to the data points of the platinum-covered (lifetime $t_{PL,Pt}$ = 5.7 ± 1.5 ns) and bare diamond side ($t_{PL,bare}$ = 13.8 ± 0.7 ns). **(b)** PL lifetimes of the five diamond samples as a function of implantation energy. **(c)** Relative lifetime (lifetime on the Pt-covered side divided by the one on the bare reference, see (b)) as a function of mean implantation depth. The data is fitted by Equation (2) with fixed exponents $p$ = 4 and $p$ = 6 and a characteristic length scale $d_0$ = 30 ± 2 nm which is found for both parameters $p$.

### 3.3 Spin properties

We further characterized the NV quantum sensing properties for the 20 keV sample, which corresponds to the lowest implantation depth (~ 27 nm) that still allows for ODMR measurements. Figure 4 (a) compares the total measurable PL from the platinum-covered with the bare diamond surface, detected with our NV-ODMR setup (Experimental Section). As in our previous measurements (Figure 2 and 3), we observe a strong reduction of the PL on the Pt-covered side. Figure 4 (b) depicts the maximally achievable Rabi contrast for both the platinum-covered and bare reference surface. The Rabi contrast of the platinum-covered diamond is significantly reduced compared to the bare diamond surface, which aligns with the increased population of the NV⁰ state and additional quenching of the NV⁻ PL (compare Figure 2 (d) and Figure 3 (c)). It is important to note that a quantitative comparison between the Rabi contrast, the NV charge state ratio and the PL quenching is challenging, as PL background sources and excitation schemes differ between our NV-ODMR and PL setup.

Next, we compare the spin-lattice relaxation time $T_1$ of NV⁻ under the bare and platinum-covered surfaces (Figure 4 (c)). The dataset is fitted with a single-exponential decay, yielding $T_1$ relaxation times of 3.29 ± 0.12 ms for the platinum-covered surface and 2.98 ± 0.07 ms for the bare diamond surface. Finally, we determine the spin coherence time $T_2$ of the NV ensemble using the Hahn echo pulse sequence (Figure 4 (d)). The two decay curves measured on the bare and the platinum-covered side both exhibit strong oscillations corresponding to the ¹³C Larmor precession frequency (~ 330 kHz) at the applied magnetic field $B_0$ of ~ 310 Gauss. Fitting the envelope of the data with a stretched exponential decay as in Equation (1) yields $T_2$ times of 71.36 ± 3.39 μs for the platinum-covered surface and 9.29 ± 0.18 μs for the bare surface.

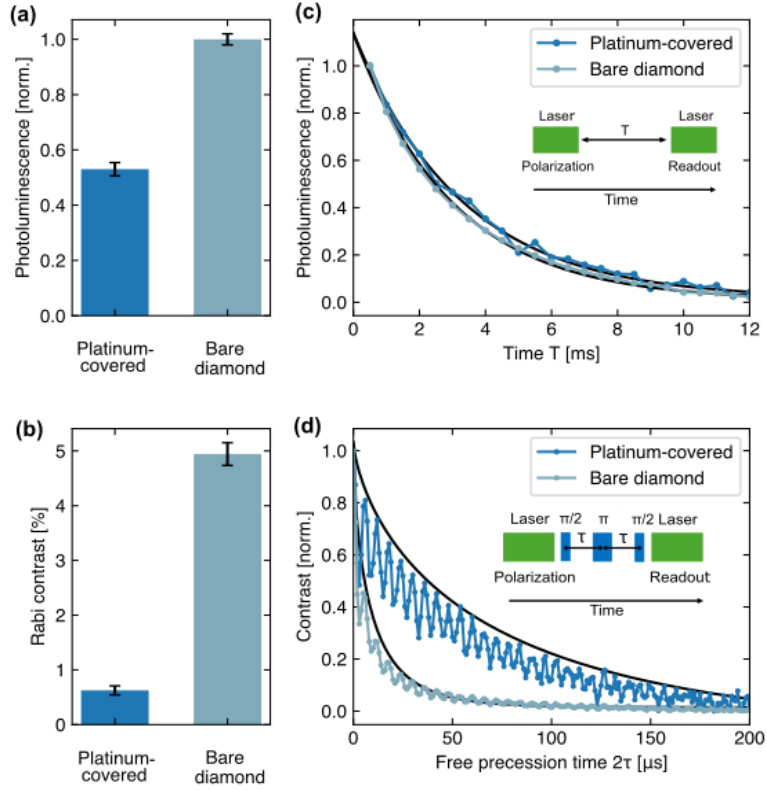

**Figure 4. NV⁻ spin properties of the 20 keV sample**. (a) Overall PL intensities of the platinum-covered and bare diamond surface observed in the 20 keV sample measured with the NV-ODMR setup. (b) Rabi contrast of the platinum-covered and bare diamond surface. (c) NV⁻ relaxation time ($T_1$) experiment for both surfaces (light and dark blue dots) together with their respective monoexponential fits (black lines). The fits result in $T_1$ of 3.29 ± 0.12 ms and 2.98 ± 0.07 ms for the platinum and reference surface, respectively. The inset depicts the pulse sequence. (d) Hahn-echo experiment (see inset for pulse sequence) to determine the coherence time $T_2$. The fits (black lines) result in coherence times of 71.36 ± 3.39 μs and 9.29 ± 0.18 μs for the platinum-covered and bare reference side, respectively.

## 4. Discussion

In the following, we analyze the observed effects of platinum thin films on near-surface NV ensembles and discuss the relationships between the distinct phenomena as well as their common origins. For shallow NV ensembles below platinum films, the comparison of the PL spectra in Figures 2 (a) and (b) shows a strong change of the emitted spectrum's shape and intensity with respect to bare diamond surfaces. This is partially caused by a redistribution of the NV population from the negative to the neutral charge state, which exhibits a generally lower PL, and an emission spectrum shifted towards lower wavelengths [72]. This observation is commonly attributed to surface-induced band bending [73–75]. Depending on the contact material and surface termination, upward bending of the diamond bands can shift the effective Fermi level below the NV⁻/NV⁰ charge state transition level and stabilize the neutral NV⁰. For strong band bending close to the surface, the Fermi level can even cross the NV⁰/NV⁺ transition level [36]. This leads to a conversion into the dark NV⁺ state, which could explain the complete PL quenching observed for the lowest implantation energies (2.5 keV and 5 keV) [76]. Additionally, charge transfer between diamond and metal or metal-induced acceptors at the interface could also shift the NV center into a less negative charge state compared to the bare diamond reference.

Besides a shift in the equilibrium charge state, additional non-radiative decay channels known as FRET, SET or MIET can occur for NV centers close to coated diamond surfaces [35,57]. FRET corresponds to a direct, resonant dipole–dipole energy transfer between spectrally overlapping emitters and absorbers, in our case the NV center and a possible point defect at the diamond–metal interface. This interaction is short ranged (~ 10 nm), yielding an exponent $p = 6$ in Equation (2) [57]. In contrast, SET and MIET describe non-resonant transfer to two-dimensional conductive surfaces. SET ($p = 4$) corresponds to dissipative transfer, e.g. via surface currents, while MIET excites surface plasmons and extends up to ~ 200 nm [35,58,63]. Figure 3 (c) shows the lifetimes ratios of the platinum-covered and bare diamond side for different implantation depths. We fit the relative reduction of lifetime to the model presented in Section 2.5 and allow for a variable order parameter $p$ to determine the characteristic length scale $d_0$ of the energy transfer. Despite a large uncertainty in the parameter $p$ that does not allow to pinpoint the transfer mechanism to FRET, SET or MIET specifically, the fit yields a characteristic length scale of $d_0 = 30 \pm 2$ nm. This interaction range points towards the dissipative SET mechanism, but the assignment should be treated with care as the fit is based only on a small number of data points.

The change of the charge state population and the reduction of NV-related PL are also reflected in the achievable Rabi contrast. These datasets are shown in Figure 4 (b) for the sample implanted with 20 keV. This sample is the shallowest sample, where a significant ODMR spectrum can be detected on the platinum-covered side. For implantation energies below 20 keV, the total NV-related PL becomes too weak to allow for measurements in the NV-ODMR set-up, but PL and Rabi contrast recover gradually to the bare reference with increasing implantation depth.

Besides the Rabi contrast and the collectable PL, also the $T_1$ and $T_2$ times are important parameters that influence the NV's sensing capabilities [12]. The $T_1$ times measured on the bare diamond side and the platinum-covered side depicted in Figure 4 (c) show no significant difference. This is surprising because other groups have observed a strong reduction of the $T_1$ time for NV$^-$ centers in the vicinity of metals [37,77,78]. This effect has been attributed to the Johnson-Nyquist noise produced by the electrons in the metal film and is directly connected to the conductivity [77,78]. We assume that this effect is less pronounced in our set of samples, because the sputtered Pt metal films studied here were significantly less conductive than the copper films in previous publications [37].

Contrary to the $T_1$ time, the $T_2$ time of the 20 keV sample changes drastically from the bare to the platinum-covered diamond side. The observed eightfold increase in the $T_2$ time on the platinum-covered side indicates a substantial suppression of noise sources within the kHz to MHz spectral range [79]. For shallow NV center ensembles, multiple factors influence the $T_2$ time such as surface-related noise [24] or the spin bath of nitrogen donors in highly doped diamonds [51,60]. A reduction of this noise can be explained by a depletion of the corresponding spin noise sources. Henshaw et al. [37] have observed this effect for shallow NV centers below copper films and attribute it to the partial ionization of near-surface nitrogen donors induced by the band bending, which also shifts the NV charge state. Alternatively, similar experiments with graphene on diamond surfaces attribute the improved coherence time to a different charge distribution at the diamond interface induced by the new layer [68,80]. Lastly, Monge et al. [32] have obtained similar results in the proximity of a superconductor and suggest that mirror charges in conductive materials can counteract fluctuations of interface spins and charge hopping and thus also contribute to the extended coherence time. More experimental and theoretical research will be needed to elucidate the underlying mechanisms.

Finally, we discuss the implications of the observed effects for quantum sensing experiments using shallow NV centers close to metallic layers. First, the reduced NV$^-$ population and the fluorescence quenching make ODMR experiments impracticable for NVs with a depth below 20 nm under our conditions. For NVs at depths of approximately 30 nm (i.e., 20 keV implant energy), the negative charge state becomes sufficiently stable, and the quenching effect sufficiently reduced, so that ODMR experiments are feasible and advanced quantum control is possible. The sensitivity for ac magnetometry in the variance detection mode, typically used for nanoscale NMR experiments, scales as

$$\eta_{ac}^{Var} \propto \frac{1}{C} \frac{1}{N_{Ph}^{1/2}} \frac{1}{T_2^{3/2}} \tag{3}$$

depending on the number of detected photons $N_{Ph}$, the achievable NV spin (Rabi) contrast $C$ and the $T_2$ time of the NV centers [37,81]. Considering only the reduced $N_{Ph}$ and $C$, we would expect a decrease in sensitivity by a factor of approximately 10. On the contrary, the coherence time $T_2$ beneath the platinum thin film is approximately eight times longer than on the bare reference side, resulting in an overall enhancement of the variance-based sensitivity $\eta_{ac}^{Var}$ by a factor of ~ 25.

However, in applications involving nanoscale spin detection, this improvement is counterbalanced by the strong dependence on the distance $r$ of the spin noise signal, which scales as $1/r^3$ for a three-dimensional ensemble of sample spins [14,16,82]. Typically, diamond samples with a nitrogen implantation energy of 2.5 keV are employed to efficiently detect spin noise of samples on the diamond surface [18,21,30]. In comparison, the NMR signal from a sample at the surface of a 20 keV-implanted diamond is approximately two orders of magnitude weaker due to the $1/r^3$ dependence. Additionally, the low gyromagnetic ratio of $^{195}$Pt further complicates spin detection, making metallic platinum NMR detection significantly more challenging than for previously reported nuclei such as $^{19}$F or $^1$H spins [18,82,83]—though not impossible. Furthermore, Henshaw et al. [37] demonstrated for copper films, that an insulating 2 nm interlayer of $Al_2O_3$ can reduce the degradation of the NV properties. In future work, a similar approach could be exploited and optimized also for sensing with platinum.

In terms of the electrical readout, our results indicate that also here the optimization of implantation parameters for NV ensembles below metal electrodes is necessary. Alternatively, the study motivates device geometries, in which NV centers and electrodes are spatially separated.

## 5. Conclusion & Outlook

Probing platinum thin films with NV centers in diamond offers numerous applications, including spintronics, catalysis, and electrochemistry [38,84]. To unlock this potential, we characterized and optimized the properties of NV centers near a platinum film deposited on a diamond chip. We showed that the properties of near-surface NV centers close to the platinum film are significantly affected: The photoluminescence quenches, the excited-state lifetimes decrease, and the charge-state ratio shifts in favor of the NV⁰ or even NV⁺ state, which are inactive for quantum sensing applications. Due to these effects, implantation energies of at least ~ 20 keV corresponding to ~ 30 nm depth are required to achieve ODMR signals beneath the platinum films. For this implantation depth, we observe a significant increase in the coherence time $T_2$ for NV centers below the platinum film compared to those on a bare diamond surface and discuss possible origins.

The presented experiments help to guide the layout of NV-based devices that either aim to sense platinum layers and their surroundings or rely on electrical readout. The observed deterioration suggests that NV centers and electrodes should be spatially separated if metallic contacts are used and shallow NV centers < 20 nm from the surface are desired. Some countermeasures against the readout deterioration that should be investigated in the future include phosphorus doping of the diamond to create a sacrificial layer or the application of an electric field to stabilize the negative charge state [29,73,85].

**Acknowledgements**

**Funding.** This study was funded by the Deutsche Forschungsgemeinschaft (DFG) within the Emmy Noether program (project id 412351169), within Germany's Excellence Strategy by the clusters MCQST (EXC-2111, 390814868) and e-conversion (EXC-2089, 390776260), as well as within the Transregios "Constrained Quantum Matter" (TRR 360, 492547816) and "Spin+X" (TRR 173, 268565370). This research is part of the Munich Quantum Valley (lighthouse project IQSense), which is supported by the Bavarian State Government with funds from the Hightech Agenda Bayern Plus. E.N. acknowledges support from the Quantum Initiative Rhineland-Palatinate (QUIP).

**Author Contributions.** D.B.B., H.H. and M.S.B. conceived and supervised the study. J.P.L. performed the experiments of the quantum properties, L.M.T. performed the photoluminescence experiments and N.K. performed the experiments for the PL lifetime. J.P.L., L.M.T. and N.K. analyzed the data. M.A. deposited the platinum layers. J.P.L., L.M.T, N.K., M.A., H.H., M.S.B, E.N. and D.B.B. discussed the results. J.P.L., L.M.T., M.S.B. and D.B.B. wrote the manuscript with input from all authors.

**Bibliography.**

1. Rovny, J. *et al.* Nanoscale diamond quantum sensors for many-body physics. *Nat. Rev. Phys.* **6**, 753–768 (2024).

2. Barry, J. F. *et al.* Sensitivity optimization for NV-diamond magnetometry. *Rev. Mod. Phys.* **92**, 015004 (2020).

3. Allert, R. D., Briegel, K. D. & Bucher, D. B. Advances in nano- and microscale NMR spectroscopy using diamond quantum sensors. *Chem. Commun.* **58**, 8165–8181 (2022).

4. Chatterjee, A. *et al.* Semiconductor qubits in practice. *Nat. Rev. Phys.* **3**, 157–177 (2021).

5. Doherty, M. W. *et al.* The nitrogen-vacancy colour centre in diamond. *Physics Reports* **528**, 1–45 (2013).

6. Schirhagl, R., Chang, K., Loretz, M. & Degen, C. L. Nitrogen-Vacancy Centers in Diamond: Nanoscale Sensors for Physics and Biology. *Annu. Rev. Phys. Chem.* **65**, 83–105 (2014).

7. Kehayias, P. *et al.* Imaging crystal stress in diamond using ensembles of nitrogen-vacancy centers. *Phys. Rev. B* **100**, 174103 (2019).

8. Kucsko, G. *et al.* Nanometre-scale thermometry in a living cell. *Nature* **500**, 54–58 (2013).

9. Bian, K. *et al.* Nanoscale electric-field imaging based on a quantum sensor and its charge-state control under ambient condition. *Nat. Commun.* **12**, 2457 (2021).

10. Le Sage, D. *et al.* Optical magnetic imaging of living cells. *Nature* **496**, 486–489 (2013).

11. Hermann, J. C. *et al.* Extending radiowave frequency detection range with dressed states of solid-state spin ensembles. *npj Quantum Information* **10**, 103 (2024).

12. Degen, C. L., Reinhard, F. & Cappellaro, P. Quantum sensing. *Rev. Mod. Phys.* **89**, 035002 (2017).

13. Levine, E. V. *et al.* Principles and techniques of the quantum diamond microscope. *Nanophotonics* **8**, 1945–1973 (2019).

14. Staudacher, T. *et al.* Nuclear Magnetic Resonance Spectroscopy on a (5-Nanometer)$^3$ Sample Volume. *Science* **339**, 561–563 (2013).

15. Sushkov, A. O. *et al.* All-Optical Sensing of a Single-Molecule Electron Spin. *Nano Lett.* **14**, 6443–6448 (2014).

16. Mamin, H. J. *et al.* Nanoscale Nuclear Magnetic Resonance with a Nitrogen-Vacancy Spin Sensor. *Science* **339**, 557–560 (2013).

17. Lovchinsky, I. *et al.* Nuclear magnetic resonance detection and spectroscopy of single proteins using quantum logic. *Science* **351**, 836–841 (2016).

18. Liu, K. S. *et al.* Surface NMR using quantum sensors in diamond. *Proc. Natl. Acad. Sci. U.S.A.* **119**, e2111607119 (2022).

19. Müller, C. *et al.* Nuclear magnetic resonance spectroscopy with single spin sensitivity. *Nat. Commun.* **5**, 4703 (2014).

20. Grafenstein, N. R. von, Briegel, K. D., Casanova, J. & Bucher, D. B. Coherent signal detection in the statistical polarization regime enables high-resolution nanoscale NMR spectroscopy. Preprint at https://doi.org/10.48550/arXiv.2501.02093 (2025).

21. Liu, K. S. *et al.* Using Metal–Organic Frameworks to Confine Liquid Samples for Nanoscale NV-NMR. *Nano Lett.* **22**, 9876–9882 (2022).

22. Kim, M. *et al.* Decoherence of Near-Surface Nitrogen-Vacancy Centers Due to Electric Field Noise. *Phys. Rev. Lett.* **115**, 087602 (2015).

23. Sangtawesin, S. *et al.* Origins of Diamond Surface Noise Probed by Correlating Single-Spin Measurements with Surface Spectroscopy. *Phys. Rev. X* **9**, 031052 (2019).

24. Dwyer, B. L. *et al.* Probing Spin Dynamics on Diamond Surfaces Using a Single Quantum Sensor. *PRX Quantum* **3**, 040328 (2022).

25. Tetienne, J.-P. *et al.* Spin properties of dense near-surface ensembles of nitrogen-vacancy centers in diamond. *Phys. Rev. B* **97**, 085402 (2018).

26. Fuhrmann, J. *et al.* Probing coherence properties of shallow implanted NV ensembles under different oxygen terminations. *Mater. Quantum. Technol.* **4**, 041001 (2024).

27. Janitz, E. *et al.* Diamond surface engineering for molecular sensing with nitrogen-vacancy centers. *J. Mater. Chem. C* **10**, 13533–13569 (2022).

28. Hauf, M. V. *et al.* Chemical control of the charge state of nitrogen-vacancy centers in diamond. *Phys. Rev. B* **83**, 081304 (2011).

29. Grotz, B. *et al.* Charge state manipulation of qubits in diamond. *Nat. Commun.* **3**, 729 (2012).

30. Rizzato, R., Bruckmaier, F., Liu, K. S., Glaser, S. J. & Bucher, D. B. Polarization Transfer from Optically Pumped Ensembles of N-V Centers to Multinuclear Spin Baths. *Phys. Rev. Applied* **17**, 024067 (2022).

31. Rizzato, R., von Grafenstein, N. R. & Bucher, D. B. Quantum sensors in diamonds for magnetic resonance spectroscopy: Current applications and future prospects. *Applied Physics Letters* **123**, 260502 (2023).

32. Monge, R. *et al.* Spin Dynamics of a Solid-State Qubit in Proximity to a Superconductor. *Nano Lett.* **23**, 422–428 (2023).

33. Tetienne, J.-P. *et al.* Quantum imaging of current flow in graphene. *Science Advances* **3**, e1602429 (2017).

34. Ziem, F., Garsi, M., Fedder, H. & Wrachtrup, J. Quantitative nanoscale MRI with a wide field of view. *Scientific Reports* **9**, 12166 (2019).

35. Tisler, J. *et al.* Single Defect Center Scanning Near-Field Optical Microscopy on Graphene. *Nano Lett.* **13**, 3152–3156 (2013).

36. Hauf, M. V. *et al.* Addressing Single Nitrogen-Vacancy Centers in Diamond with Transparent in-Plane Gate Structures. *Nano Lett.* **14**, 2359–2364 (2014).

37. Henshaw, J. *et al.* Mitigation of nitrogen vacancy photoluminescence quenching from material integration for quantum sensing. *Mater. Quantum. Technol.* **3**, 035001 (2023).

38. Lu, C. *et al.* UHV, Electrochemical NMR, and Electrochemical Studies of Platinum/Ruthenium Fuel Cell Catalysts. *J. Phys. Chem. B* **106**, 9581–9589 (2002).

39. Madima, N. & Raphulu, M. Platinum group metals-based electrodes for high-performance lithium-oxygen batteries: A mini-review. *Journal of Electroanalytical Chemistry* **976**, 118799 (2025).

40. Still, B. M., Kumar, P. G. A., Aldrich-Wright, J. R. & Price, W. S. 195Pt NMR—theory and application. *Chem. Soc. Rev.* **36**, 665–686 (2007).

41. Nakayama, H. *et al.* Spin Hall Magnetoresistance Induced by a Nonequilibrium Proximity Effect. *Phys. Rev. Lett.* **110**, 206601 (2013).

42. Sinova, J., Valenzuela, S. O., Wunderlich, J., Back, C. H. & Jungwirth, T. Spin Hall effects. *Rev. Mod. Phys.* **87**, 1213–1260 (2015).

43. Althammer, M. *et al.* Quantitative study of the spin Hall magnetoresistance in ferromagnetic insulator/normal metal hybrids. *Phys. Rev. B* **87**, 224401 (2013).

44. Cornelissen, L. J., Liu, J., Duine, R. A., Youssef, J. B. & van Wees, B. J. Long-distance transport of magnon spin information in a magnetic insulator at room temperature. *Nature Phys.* **11**, 1022–1026 (2015).

45. Bourgeois, E. *et al.* Photoelectric detection of electron spin resonance of nitrogen-vacancy centres in diamond. *Nat. Commun.* **6**, 8577 (2015).

46. Hrubesch, F. M., Braunbeck, G., Stutzmann, M., Reinhard, F. & Brandt, M. S. Efficient Electrical Spin Readout of NV-Centers in Diamond. *Phys. Rev. Lett.* **118**, 037601 (2017).

47. Bourgeois, E., Gulka, M. & Nesladek, M. Photoelectric Detection and Quantum Readout of Nitrogen-Vacancy Center Spin States in Diamond. *Advanced Optical Materials* **8**, 1902132 (2020).

48. Bucher, D. B. *et al.* Quantum diamond spectrometer for nanoscale NMR and ESR spectroscopy. *Nat. Protoc.* **14**, 2707–2747 (2019).

49. Ziegler, J. F., Ziegler, M. D. & Biersack, J. P. SRIM – The stopping and range of ions in matter (2010). *Nuclear Instruments and Methods in Physics Research Section B: Beam Interactions with Materials and Atoms* **268**, 1818–1823 (2010).

50. Pham, L. M. *et al.* NMR technique for determining the depth of shallow nitrogen-vacancy centers in diamond. *Phys. Rev. B* **93**, 045425 (2016).

51. Henshaw, J. *et al.* Nanoscale solid-state nuclear quadrupole resonance spectroscopy using depth-optimized nitrogen-vacancy ensembles in diamond. *Appl. Phys. Lett.* **120**, 174002 (2022).

52. Flacke, L. *et al.* Robust formation of nanoscale magnetic skyrmions in easy-plane anisotropy thin film multilayers with low damping. *Phys. Rev. B* **104**, L100417 (2021).

53. Schlitz, R. *et al.* Electrically Induced Angular Momentum Flow between Separated Ferromagnets. *Phys. Rev. Lett.* **132**, 256701 (2024).

54. Davies, G. Vibronic spectra in diamond. *J. Phys. C: Solid State Phys.* **7**, 3797–3809 (1974).

55. Su, Z. *et al.* Luminescence landscapes of nitrogen-vacancy centers in diamond: quasi-localized vibrational resonances and selective coupling. *J. Mater. Chem. C* **7**, 8086–8091 (2019).

56. Waldherr, G. *et al.* Dark States of Single Nitrogen-Vacancy Centers in Diamond Unraveled by Single Shot NMR. *Phys. Rev. Lett.* **106**, 157601 (2011).

57. Förster, Th. Zwischenmolekulare Energiewanderung und Fluoreszenz. *Annalen der Physik* **437**, 55–75 (1948).

58. Gregor, I., Chizhik, A., Karedla, N. & Enderlein, J. Metal-induced energy transfer. *Nanophotonics* **8**, 1689–1699 (2019).

59. Freire-Moschovitis, F. A. *et al.* The Role of Electrolytes in the Relaxation of Near-Surface Spin Defects in Diamond. *ACS Nano* **17**, 10474–10485 (2023).

60. Bauch, E. *et al.* Decoherence of ensembles of nitrogen-vacancy centers in diamond. *Phys. Rev. B* **102**, 134210 (2020).

61. Pezzagna, S., Naydenov, B., Jelezko, F., Wrachtrup, J. & Meijer, J. Creation efficiency of nitrogen-vacancy centres in diamond. *New J. Phys.* **12**, 065017 (2010).

62. Yun, C. S. *et al.* Nanometal Surface Energy Transfer in Optical Rulers, Breaking the FRET Barrier. *J. Am. Chem. Soc.* **127**, 3115–3119 (2005).

63. Nelz, R. *et al.* Near-Field Energy Transfer between a Luminescent 2D Material and Color Centers in Diamond. *Advanced Quantum Technologies* **3**, 1900088 (2020).

64. Maletinsky, P. *et al.* A robust scanning diamond sensor for nanoscale imaging with single nitrogen-vacancy centres. *Nature Nanotech.* **7**, 320–324 (2012).

65. Israelsen, N. M. *et al.* Increasing the photon collection rate from a single NV center with a silver mirror. *J. Opt.* **16**, 114017 (2014).

66. Lillie, S. E. *et al.* Magnetic noise from ultrathin abrasively deposited materials on diamond. *Phys. Rev. Materials* **2**, 116002 (2018).

67. Li, D.-F. *et al.* Thickness dependent surface plasmon of silver film detected by nitrogen vacancy centers in diamond. *Opt. Lett.* **43**, 5587 (2018).

68. Hao, Y. *et al.* Coherence enhancement via a diamond-graphene hybrid for nanoscale quantum sensing. *National Science Review* **12**, nwaf076 (2025).

69. Storteboom, J., Dolan, P., Castelletto, S., Li, X. & Gu, M. Lifetime investigation of single nitrogen vacancy centres in nanodiamonds. *Opt. Express* **23**, 11327 (2015).

70. Radko, I. P. *et al.* Determining the internal quantum efficiency of shallow-implanted nitrogen-vacancy defects in bulk diamond. *Opt. Express* **24**, 27715 (2016).

71. Fuchs, P., Challier, M. & Neu, E. Optimized single-crystal diamond scanning probes for high sensitivity magnetometry. *New J. Phys.* **20**, 125001 (2018).

72. Beha, K., Batalov, A., Manson, N. B., Bratschitsch, R. & Leitenstorfer, A. Optimum Photoluminescence Excitation and Recharging Cycle of Single Nitrogen-Vacancy Centers in Ultrapure Diamond. *Phys. Rev. Lett.* **109**, 097404 (2012).

73. Hauf, M. Charge state control of nitrogen-vacancy centers in diamond. (PhD thesis, Technische Universität München, München, 2013).

74. Zhang, Z. & Yates, J. T. Band Bending in Semiconductors: Chemical and Physical Consequences at Surfaces and Interfaces. *Chem. Rev.* **112**, 5520–5551 (2012).

75. Broadway, D. A. *et al.* Spatial mapping of band bending in semiconductor devices using in situ quantum sensors. *Nat. Electron.* **1**, 502–507 (2018).

76. Dickmann, M. *et al.* Identification and Reversible Optical Switching of $NV^+$ Centers in Diamond. *Advanced Functional Materials* **35**, 2500817 (2025).

77. Ariyaratne, A., Bluvstein, D., Myers, B. A. & Jayich, A. C. B. Nanoscale electrical conductivity imaging using a nitrogen-vacancy center in diamond. *Nat. Commun.* **9**, 2406 (2018).

78. Kolkowitz, S. *et al.* Probing Johnson noise and ballistic transport in normal metals with a single-spin qubit. *Science* **347**, 1129–1132 (2015).

79. Han, S. *et al.* Solid-state spin coherence time approaching the physical limit. *Sci. Adv.* **11**, eadr9298 (2025).

80. Lo, W. K. *et al.* Enhancement of quantum coherence in solid-state qubits via interface engineering. *Nat. Commun.* **16**, 5984 (2025).

81. Du, J., Shi, F., Kong, X., Jelezko, F. & Wrachtrup, J. Single-molecule scale magnetic resonance spectroscopy using quantum diamond sensors. *Rev. Mod. Phys.* **96**, 025001 (2024).

82. Kehayias, P. *et al.* Solution nuclear magnetic resonance spectroscopy on a nanostructured diamond chip. *Nat. Commun.* **8**, 188 (2017).

83. Staudacher, T. *et al.* Probing molecular dynamics at the nanoscale via an individual paramagnetic centre. *Nat. Commun.* **6**, 8527 (2015).

84. Hirohata, A. *et al.* Review on spintronics: Principles and device applications. *Journal of Magnetism and Magnetic Materials* **509**, 166711 (2020).

85. Watanabe, A. *et al.* Shallow NV centers augmented by exploiting n-type diamond. *Carbon* **178**, 294–300 (2021).